\documentclass[final]{svjour2}
\usepackage{graphicx}
\usepackage{subfigure}
\usepackage{rotating}
\usepackage{amssymb}
\usepackage{mathptmx}
\usepackage[numbers]{natbib}
\makeatletter
\journalname{Journal of Low Temperature Physics}
\bibpunct{}{}{,}{s}{}{,}

\begin{document}

\newcommand{\hdblarrow}{H\makebox[0.9ex][l]{$\downdownarrows$}-}
\title{Simulations of Noise in Phase-Separated Transition-Edge Sensors for SuperCDMS}

\author{A.J. Anderson$^1$ \and S.W. Leman$^1$ \and M. Pyle $^2$ \and E. Figueroa-Feliciano$^1$ \and K. McCarthy$^1$ \and T. Doughty$^3$ \and M. Cherry$^2$ \and B. Young$^4$ \and for the SuperCDMS Collaboration}

\institute{1:Department of Physics, Massachusetts Institute of Technology, Cambridge, MA 02139, USA\\
\email{adama@mit.edu}\\
2:Department of Physics, Stanford University, Stanford, CA 94305, USA\\
3:Department of Physics, University of California, Berkeley, CA 94720, USA\\
4:Department of Physics, Santa Clara University, Santa Clara, CA 95053, USA\\}

\date{\today}

\maketitle


\begin{abstract}

We briefly review a simple model of superconducting-normal phase-separation in transition-edge sensors (TES) in the SuperCDMS experiment.  After discussing some design considerations relevant to the TES in the detectors, we study noise sources in both the phase-separated and phase-uniform cases.  Such simulations will be valuable for optimizing the critical temperature and TES length of future SuperCDMS detectors.
\end{abstract}

\section{Introduction}
When thermal diffusion along the length of a long, narrow transition-edge sensor (TES) is weak, a thermal gradient tends to develop, causing the device to separate into a region in the superconducting phase and a region in the normal phase.  In applications such as detectors used in the SuperCDMS experiment, long TES with high resistance are required, and this normal-superconducting phase separation is a relevant effect.  Phase separation has important consequences for detector design because it modifies the TES dynamics and affects the noise of the device.

CDMS detectors use Al fins patterned on a Ge substrate to collect athermal prompt phonons, and this energy is transported by quasiparticles into a parallel array of W TES.  The measurement of athermal phonons allows the position of an event to be inferred from the partition of energy between different channels on the detector surface.  The signal can be increased by increasing the surface coverage of the Al, but enhancing the Al coverage also requires increasing the number of TES on the surface.  However, adding TES in parallel reduces the resistance of each readout channel, and therefore reduces the bandwidth.  Because the bandwidth must be maintained large enough to resolve the signal, CDMS maximizes resistance by using TES that have the largest possible length-to-width ratio.\cite{QET}.

TES that are long enough to produce phase separation may have additional noise resulting from fluctuations of the phase boundary location along the length of the TES.  Given that the increased Al coverage resulting from longer TES increases the signal, it is important to understand how phase separation in long TES increases the noise.  We implement a simple simulation of several noise sources in the detectors, including noise from internal thermal fluctuations, and discuss how noise increases in phase-separated devices.

\section{Simulation and Sources of Noise}
We consider a simple model in which each TES is divided into a one-dimensional set of discrete nodes, each of which has its own temperature and resistance.  A one-dimensional simulation is a good approximation to CDMS TES because of their large length-to-width ratio ($\sim$100:1).  The TES nodes are each thermally coupled to a bath, and they are also thermally coupled to each other by a discrete diffusion equation\cite{CabreraTESPhysics}\cite{SteveThesis}.

\begin{figure}[ht]
\begin{center}
\includegraphics[scale=0.7]{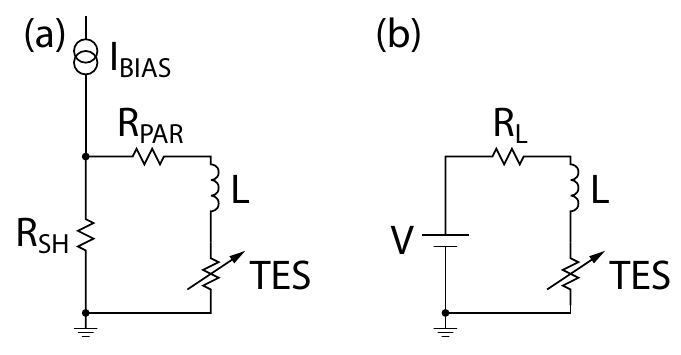}
\caption{Basic TES circuit diagrams\cite{IrwinHilton}: a.) shows the true circuit, and b.) shows the Thevanin equivalent circuit.\label{fig:ITFN}}
\end{center}
\end{figure}

The physical parameters describing the CDMS TES are difficult to determine \emph{a priori} because of variability from fabrication, and the Al fins that attach to the TES also modify the effective heat capacity and electron-phonon coupling between the TES and Ge substrate.    A procedure has been developed to estimate these model parameters by performing a least-squares fit of data to simulated I-V curves and the simulated current response to a square bias wave, as described in more detail elsewhere\cite{TESTuning}.  With this method, the main model parameters--normal resistivity ($\rho_n$), parasitic resistance ($R_{par}$), effective electron-phonon coupling ($\Sigma$), effective specific heat capacity ($c_p$), and thermal diffusion constant ($D$)--can be estimated, albeit with some uncertainty.  The critical temperature $T_c$ and the 10-90\% transition width $\Delta T_{10-90}$ were determined from critical current measurements and resistance measurements.  For SuperCDMS detectors, the TES length is $\ell = 220$~$\mu$m.

\begin{table}
\begin{center}
\begin{tabular}{l | c | c}
Parameter & Phase-separated & Phase-Uniform\\ \hline
$T_c$ & 90~mK & 80~mK \\
$\Delta T_{1090}$ & 0.5~mK & 0.6~mK \\
$\Sigma$ & $4.8 \times 10^8$~W~m$^{-3}$~K$^{-5}$ & $4.8 \times 10^8$~W~m$^{-3}$~K$^{-5}$  \\
$\rho_n$ & $1.2 \times 10^{-7}$~$\Omega$~m & $1.2 \times 10^{-7}$~$\Omega$~m \\
$R_{par}$ & 15~m$\Omega$ & 15~m$\Omega$\\
$R_{sh}$ & 20~m$\Omega$ & 20~m$\Omega$\\
$D$ & $4.03 \times 10^{-4}$~m$^2$~s$^{-1}$ & $3.77 \times 10^{-4}$~m$^2$~s$^{-1}$ \\
$c_p$ & $37.0$~J~K$^{-1}$ m$^{-3}$ & $39.7$~J~K$^{-1}$~m$^{-3}$ \\
$\ell$ & 220~$\mu$m & 140~$\mu$m \\
$\alpha$ & 500 & 187 \\
$N$ & 16 & 16 \\
$\Delta t$ & $8 \times 10^{-8}$~s & $8 \times 10^{-8}$~s
\end{tabular}
\caption{Characteristic model parameters found from matching of I-V curves and square-wave response.\label{tab:characteristicParams}}
\end{center}
\end{table}

There are several primary sources of noise in this model that we consider, which are common to many TES devices\cite{IrwinHilton}:

\begin{itemize}
\item \emph{Johnson noise from shunt resistor and parasitic resistance}: The shunt resistor in parallel with the TES and the parasitic resistance in the circuit produce Johnson noise with a conventional white spectrum
\begin{equation}
\delta V_{sh} = \sqrt{4 k_B T_{sh}R_{sh} + 4k_B T_p R_p} ,
\end{equation}
where $T_{sh}$ and $T_p$ are the temperatures of the shunt resistor and parasitic resistance, and $R_{sh}$ and $R_p$ are the values of the shunt resistor and parasitic resistance.

\item \emph{Johnson noise from TES}: Each node of the TES produces its own Johnson noise also with the standard white spectrum
\begin{equation}
\delta V_i = \sqrt{4 k_B T_i \rho(T_i) \Delta L / A_{cs}},
\end{equation}
where $\rho(T_i)$ is the resistivity of the $i$th node in the TES, $\Delta L = \ell / N$ is the length of each node, and $A_{cs}$ is the cross-sectional area.  The key distinction between Johnson noise from the TES and the Johnson noise from the other terms is that there is a correlation between voltage fluctuations in the TES and thermal fluctuations due to Joule heating.  This correlation leads to a suppression of the TES Johnson noise at low frequencies, relative to the noise from the shunt and parasitic resistances.  

\item \emph{Phonon noise to bath}: Each node of the TES is thermally coupled to the substrate by a conductance $G_{ep}^i = 5 \Sigma V_i T_i^4$, where $\Sigma$ is a constant equal to about $4 \times 10^8$ Wm$^{-3}$K$^{-5}$, $V_i$ is the volume of the $i$th node, and $T_i$ is the temperature of the $i$th node.  The spectral density of thermal fluctuations to the bath is white with variance
\begin{equation}
\delta P_{ep} = \sqrt{2k_B (T_i^2 + T_{bath}^2) G^i_{ep}}.
\end{equation}

\item \emph{Internal thermal fluctuation noise (ITFN)}: We allow power to be transferred between adjacent nodes of the TES.  The spectrum of these thermal power fluctuations is the same as for thermal fluctuations to the bath, except that the conductivity is the thermal conductivity of the TES material, W in our case.  If we assume that the conductivity is a value $g_{wf}$ consistent with the Wiedemann-Franz law, then the power spectrum of thermal fluctuations between adjacent nodes is given by
\begin{equation}
\delta P_{i,i+1} = \sqrt{4k_B \left( \frac{T_i + T_{i+1}}{2} \right)^2 G_{wf} \frac{A_{cs}}{\ell / N}},
\end{equation}
where $N$ is the number of divisions of the TES. 
\end{itemize}

The noise is added to a simulation of the circuit elements  with discrete steps in the time domain and then converted to current noise spectra.  Simulations were run with 16 nodes in each TES.  All noise sources were checked for dependence on $N$, and the average current noise was found to be independent of the number of nodes for $N$ between at least 4 and 24, though statistical fluctuations were smaller with a larger number of nodes.

\section{Results and Discussion}
We consider the noise simulation in two cases: a phase-uniform TES and a phase-separated TES.  The phase-uniform setting is valuable because analytic noise spectra are well-understood and can be calculated, allowing us to assess the validity of the simulation.  The phase-separated case is interesting because it is less analytically tractable, and because current SuperCDMS detectors exhibit phase-separation.  We can therefore compare our phase-separated simulation to CDMS data to better understand the various components of the noise.

\begin{figure}[ht]
\centering
\subfigure[Simulation]{
\includegraphics[scale=0.28]{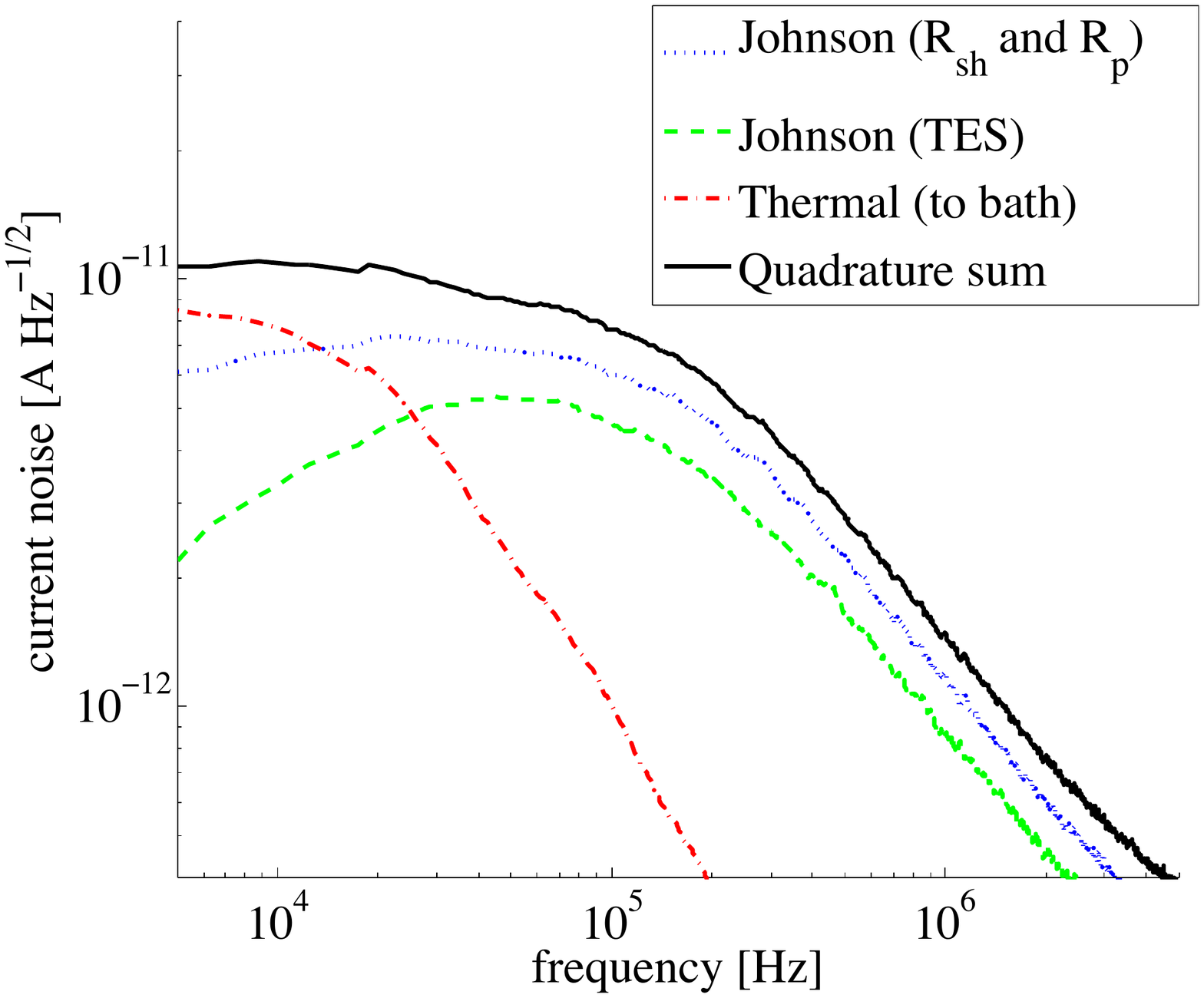}
\label{fig:PhaseUniSim}
}
\subfigure[Analytic model]{
\includegraphics[scale=0.28]{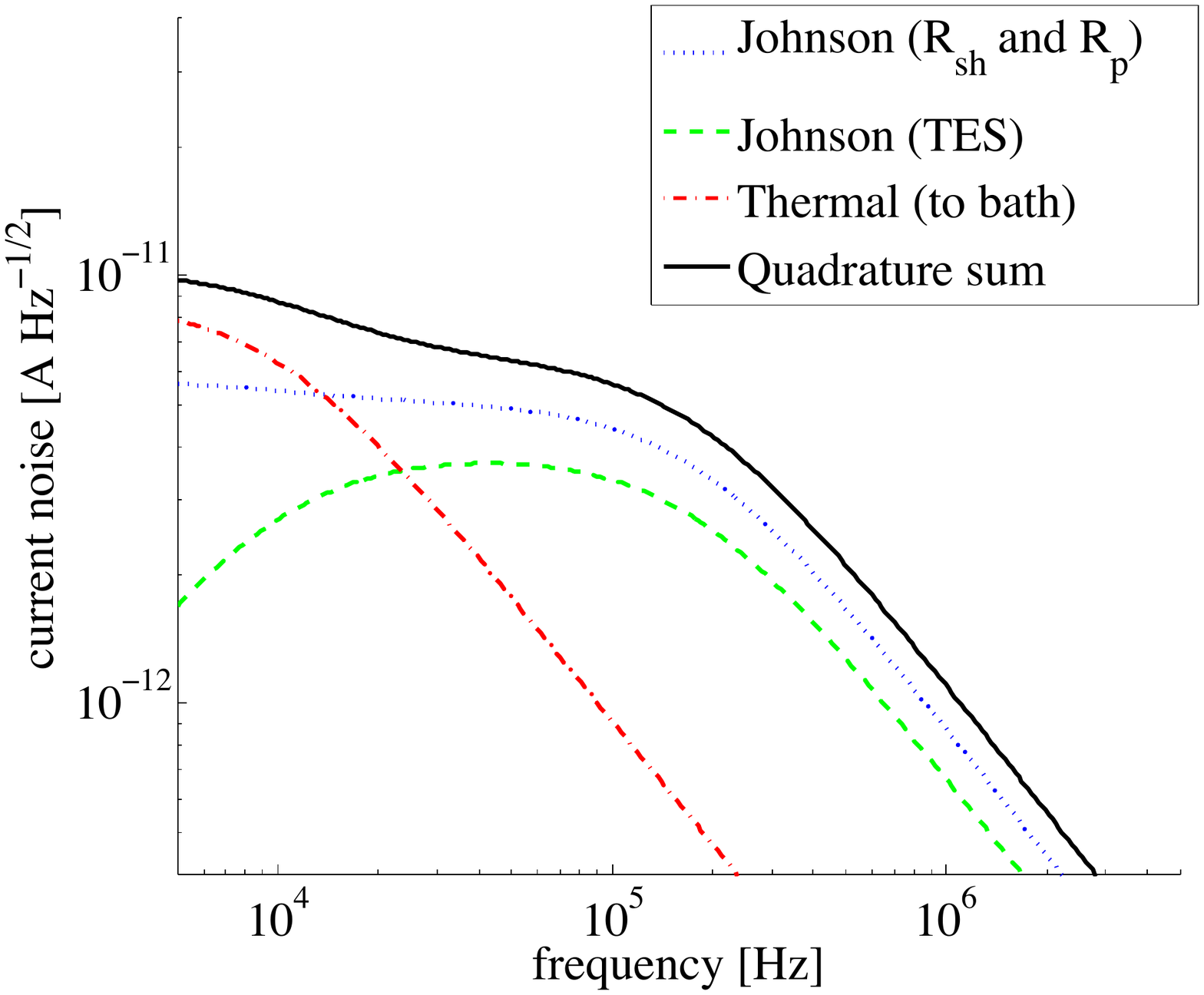}
\label{fig:PhaseUniTheory}
}
\label{fig:subfigureExample}
\caption[Optional caption for list of figures]{Current noise spectra for a homogeneous phase-uniform TES with $T_c = 80$~mK and $\ell = 140$~$\mu$m.  Different lines show the contributions due to the TES Johnson noise, the Johnson noise of the shunt resistor and parasitic resistance, and thermal noise between the TES and bath.}
\end{figure}

Figures \ref{fig:PhaseUniSim} and \ref{fig:PhaseUniTheory} show the noise from a homogeneous phase-uniform TES for the analytic model and the simulation, respectively.  The agreement between the two is qualitatively good.  At high frequencies, the Johnson noise rolls off due to the $L/R$ time constant of the circuit.  At lower frequencies the thermal noise to the bath rolls off due to time constant for electrothermal feedback, and the TES Johnson noise shows the suppression due to the correlation with Joule heating discussed earlier.

\begin{figure}[ht]
\centering
\subfigure[Simulation]{
\includegraphics[scale=0.28]{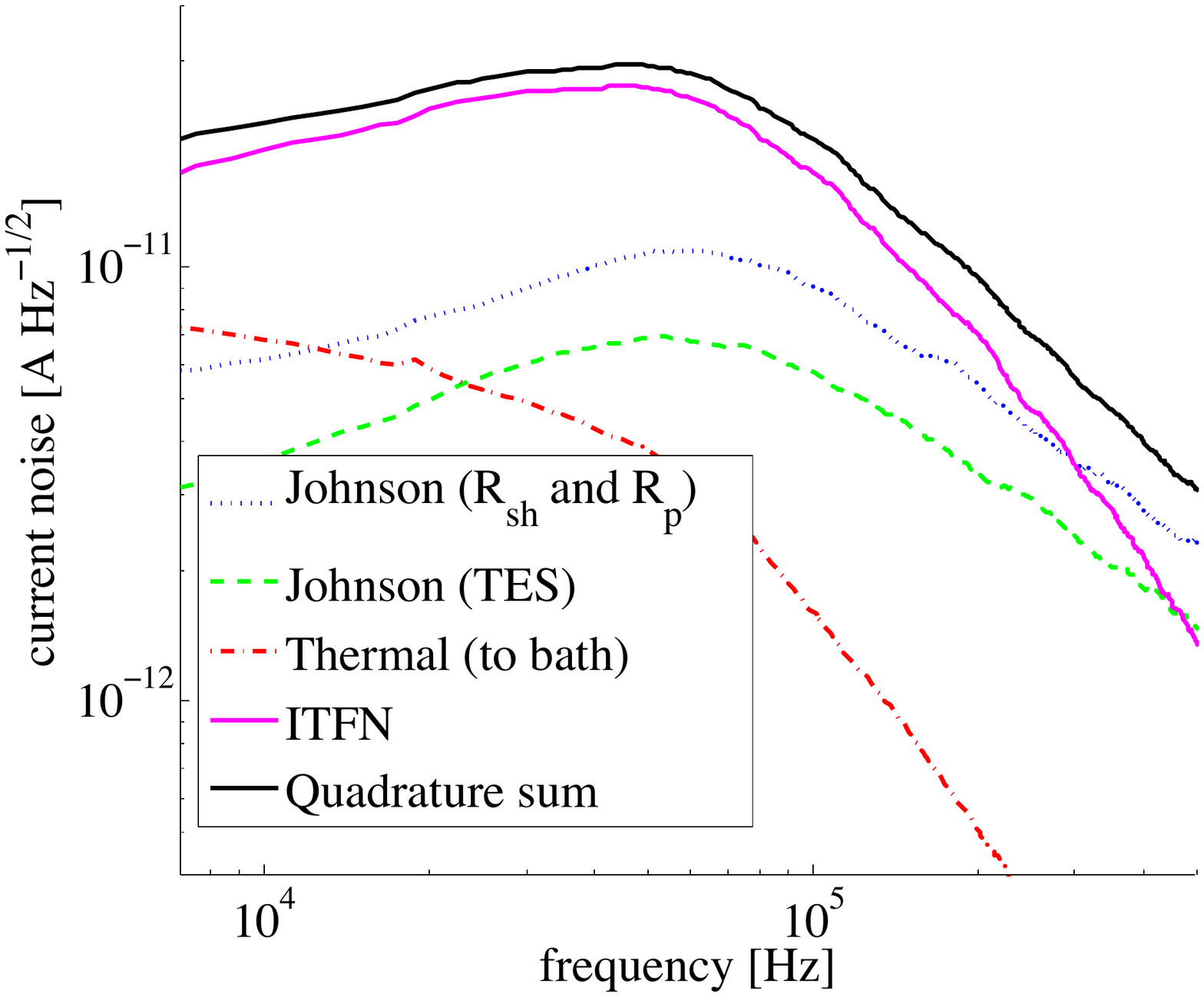}
\label{fig:PhaseSepSim}
}
\subfigure[Data from several channels of a CDMS detector]{
\includegraphics[scale=0.28]{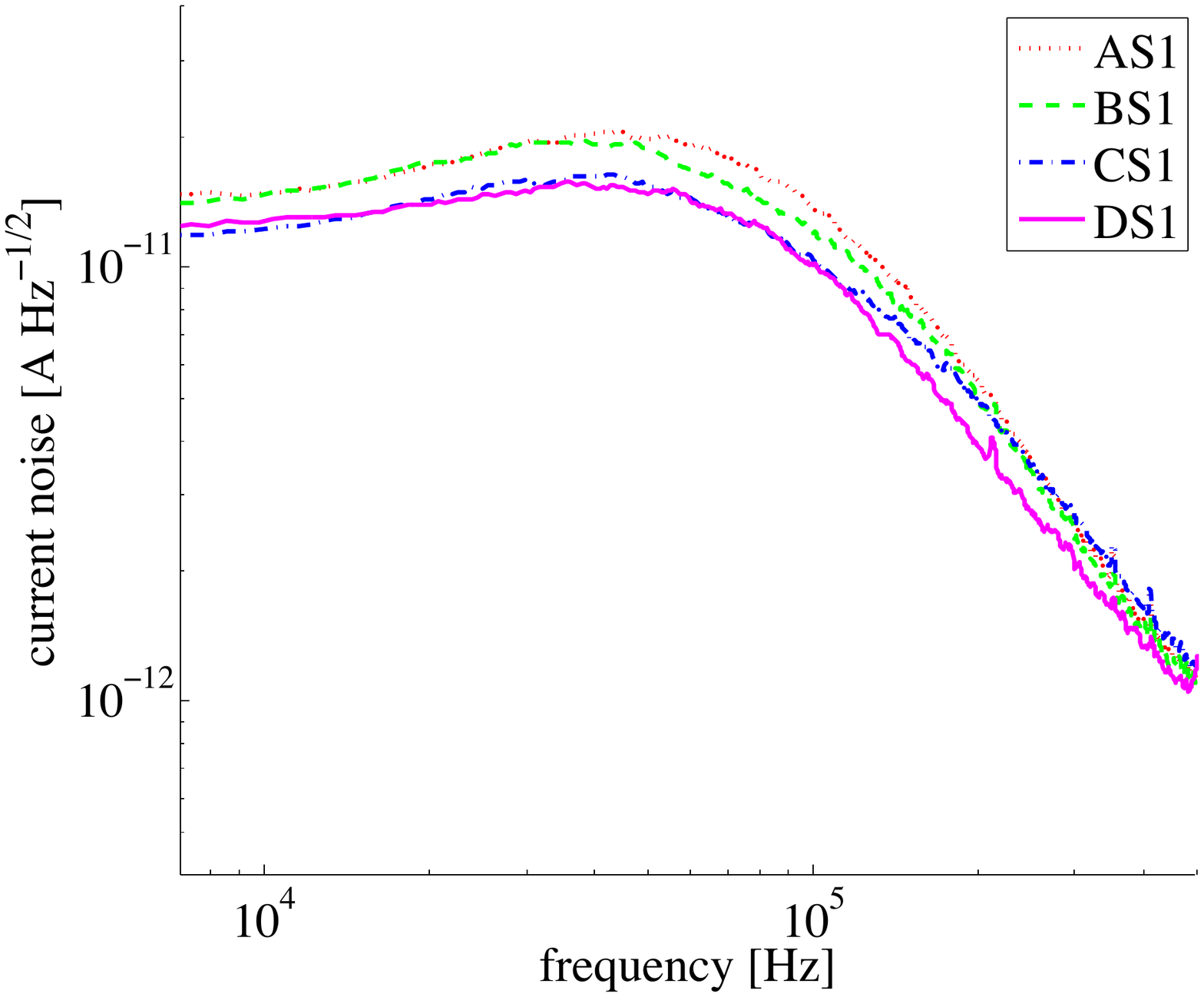}
\label{fig:PhaseSepTheory}
}
\label{fig:subfigureExample}
\caption[Optional caption for list of figures]{Current noise spectra for phase-separated TES with $T_c = 91$~mK and $\ell = 220$~$\mu$m, typical values for the detector from which the noise spectrum is taken.}
\end{figure}

Increasing the length of the TES and the critical temperature, we produce phase-separation in the TES.  With these new parameters, Figures \ref{fig:PhaseSepSim} and \ref{fig:PhaseSepTheory} show the new current noise spectra, compared with test data from CDMS detectors operating at the Soudan mine.  There is a significant enhancement in the simulated noise around 50~kHz due primarily to the ITFN.  The enhancement of the ITFN when phase-separated is made explicit in Figure \ref{fig:ITFN}, which shows that this noise source increases by about an order of magnitude.  Interestingly, the noise from thermal fluctuations to the bath and Johnson noise terms show very little change between phase-separated and phase-uniform configurations.  Overall, the simulated noise is slightly larger than measured in the data, but the discrepancy is not excessive.

\begin{figure}[ht]
\begin{center}
\includegraphics[scale=0.4]{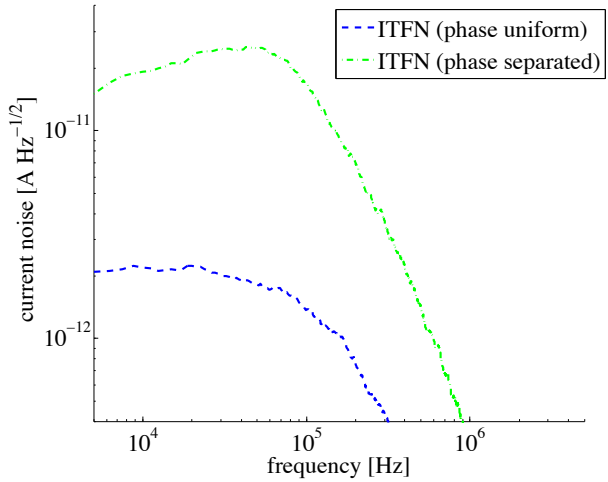}
\caption{Comparison of simulated ITFN for phase uniform and phase separated TES, showing a clear enhancement in noise. \label{fig:ITFN}}
\end{center}
\end{figure}

Other sources of noise contribute to the detector, such as SQUID and amplifier noise for example, but we do not include these in the simulation.

\section{Conclusions and Future Work}
We have implemented a numerical simulation of several important noise sources in CDMS TES.  The simulation appears to agree well with an analytic model in the phase uniform case, and the agreement with the experimental data is reasonable.  On the one hand, the inclusion of noise is a useful addition to the CDMS detector Monte Carlo simulation.  On the other hand, the results confirm previous investigations that suggest that the current noise in CDMS can be significantly reduced by eliminating phase separation.  Indeed, figure \ref{fig:PhaseSepSim} suggests that nearly half of the noise in current detectors is due to internal thermal fluctuations associated with phase-separation.  The noise could therefore be improved significantly by upgrading the SQUIDs used in readout, or by using phase-uniform TES.  Phase-uniformity could be achieved either through lower critical temperatures or shorter devices.

Finally, a working simulation will allow us to carry out a more careful optimization of the TES parameters and noise.  In particular, it will be possible to understand more quantitatively the trade-off between the improved Al surface coverage and the worse noise from phase separation.

\begin{acknowledgements}
This work is supported by the United States National Science Foundation under Grant No. PHY-0847342.
\end{acknowledgements}

\end{document}